%% file: 0_main.tex
\documentclass[12pt]{article}
\usepackage{amsmath,cite,epsfig,a4,times,euscript,listings}
\usepackage[text-hat-accent]{euler}
\usepackage{graphics}
\usepackage[usenames]{color}
\renewcommand{\baselinestretch}{1.1}
\newcommand{\myTitle}[1]{\begin{center}{\bf\Huge #1}\\[5ex]\end{center}}
\newcommand{\myAuthor}[1]{\begin{center}{\Large #1}\\[2ex]\end{center}}
\newcommand{\myAffiliation}[1]{\\[1ex]{\it\large #1}}
\newcommand{\myEmail}[1]{}
\newcommand{\myDate}{\begin{center}{\large\today}\\[5ex]\end{center}}
\newcommand{\myAbstract}[1]{\begin{center}\renewcommand{\baselinestretch}{1}{\bf Abstract}\\[2ex]\parbox{0.8\linewidth}{\small\hspace{15pt} #1}\end{center}\vspace{\baselineskip}}
\newcommand{\myReport}[1]{\hspace{\fill} #1}
\newcommand{\myPreprint}[1]{}
\newcommand{\myKeywords}[1]{}

\newcommand{\myScript}[1]{\EuScript{#1}}

\newcommand{\fudgeb}{\\[-0.7ex]}

\newcommand{\slashp}{p\hspace{-6.5pt}/}

\newcommand{\slashk}{k\hspace{-6.5pt}/}

\newcommand{\Appendix}[1]{Appendix~\ref{#1}}   
\newcommand{\Section}[1]{Section~\ref{#1}}

\newcommand{\Equation}[1]{Eq.~(\ref{#1})}

\newcommand{\eg}{{\it e.g.}}
\newcommand{\Amp}{\myScript{A}}
\newcommand{\srac}[2]{{\textstyle\frac{#1}{#2}}}
\newcommand{\Tr}{\mathrm{Tr}}

\newcommand{\vep}{\varepsilon}

\newcommand{\ANG}[1]{\langle#1\rangle}
\newcommand{\SQR}[1]{[#1]}
\newcommand{\AxS}[3]{\langle#1|#2|#3]}

\newcommand{\SxS}[3]{[#1|#2|#3]}

\newcommand{\RA}[1]{|#1\rangle}

\newcommand{\RS}[1]{|#1]}
\newcommand{\lop}[2]{#1\!\cdot\!#2}
\newcommand{\kapp}{\kappa}

\newcommand{\kstr}{\kappa^*}

\newcommand{\codetxt}[1]{{\small\tt #1}}
\newcommand{\tweakcodepar}[3]%
  {\vspace{#1ex}\newline\noindent\hspace*{4.0ex}{\small\tt #3}\vspace{#2ex}\newline\noindent}

\begin{document}

\myReport{IFJPAN-IV-2015-4}
\myPreprint{}\\[2ex]

\myTitle{%
Numerical evaluation of\\[-0.5ex] multi-gluon amplitudes for\\[0.5ex] High Energy Factorization
}

\myAuthor{%
M.~Bury and A.~van~Hameren%
\myAffiliation{%
The H.\ Niewodnicza\'nski Institute of Nuclear Physics\fudgeb
Polisch Academy of Sciences\\
Radzikowskiego 152, 31-342 Cracow, Poland%
\myEmail{hameren@ifj.edu.pl}
}
}

\myDate

\myAbstract{%
We present a program to evaluate tree-level multi-gluon amplitudes with up to two of them off-shell.
Furthermore, it evaluates squared amplitudes summed over colors and helicities for up to six external gluons.
It employs both analytic expressions, obtained via BCFW recursion, and numerical BCFW recursion.
It has been validated numerically with the help of an independent program employing numerical Dyson-Schwinger recursion.
}

\myKeywords{QCD}

%

\input{1_intro.tex}

\input{2_definitions.tex}

\input{3_color.tex}
\input{4_usage.tex}
\section{\label{Sec:summary}Summary}
We presented a program to numerically evaluate multi-gluon amplitudes, which can be applied in factorized cross section calculations for hadron collisions which ask for off-shell initial-state gluons.
It calculates squared amplitudes summed over colors and helicities for up to six external gluons with up to two of them off-shell.
Analytical expressions for color ordered helicity amplitudes are employed for up to five external gluons with up to two of them off-shell.
Higher multiplicities are evaluated using numerical BCFW recursion.
The program has been validated with an independent library employing numerical Dyson-Schwinger recursion.

\subsection*{Acknowledgments}
The authors would like to thank K.~Kutak for useful comments and suggestions.
M.B.\ was supported by Narodowe Centrum Nauki with Sonata Bis grant DEC-2013/10/E/ST2/00656. 
A.v.H.\ was supported by the Research Funding Program ARISTEIA, HOCTools
(co-financed by the European Union (European Social Fund ESF) and Greek national
funds through the Operational Program "Education and Lifelong Learning" of the National
Strategic Reference Framework (NSRF)).

\input{8_references.tex}

\begin{appendix}
\input{9_appendixA.tex}
\end{appendix}

\end{document}

%% file: 1_intro.tex
\section{Introduction\label{Sec:intro}}
Factorization prescriptions are powerful tools to tame the complex calculations involving quantum chromodynamics (QCD) for scattering processes at collider experiments like at the Large Hadron Collider.
They factorize contributions to cross sections according to the scales involved, and/or according to universality and accessibility via perturbation theory.
Many factorization prescriptions are heuristic in nature, and some are proven, which means that perturbation theory and the treatment of possible singularities, along with the occurrence of large logarithms of ratios of scales, can be dealt with in a systematic manner.

Partonic scattering amplitudes form an essential ingredient in factorized calculations of cross sections for hadron collisions.
Even though the partonic states are not physical, on-shell partonic amplitudes are well defined gauge invariant objects of the gauge theory QCD, using its Lagrangian and the Lehmann Symanzik Zimmermann reduction formula.
Factorization embeds the non-physical scattering amplitudes into physical cross sections.
Recently, it has been shown that scattering amplitudes involving any number of {\em off-shell\/} external gluons can also be defined in a rigorous manner~\cite{Kotko:2014aba}.
Such amplitudes are relevant in factorization prescriptions requiring off-shell initial-state partons, like High Energy Factorization (HEF)~\cite{Collins:1991ty,Catani:1990eg}.
Recent developments and calculations involving such factorization prescriptions can be found in ~\cite{%
Kutak:2012rf%
,Maciula:2013kd%
,Nefedov:2013ywa%
,Kutak:2014wga%
,Dooling:2014kia%
,Lipatov:2014yna%
,vanHameren:2014ala%
,Cruz-Santiago:2015nxa%
,Kotko:2015ura%
}

Calculations employing collinear factorization, for which the scattering amplitudes are completely on-shell, have been automated to the end for arbitrary processes, with essentially arbitrary multiplicities, and within essentially arbitrary models of quantum field theory~\cite{%
 Boos:1994xb
,Stelzer:1994ta
,Maltoni:2002qb
,Mangano:2002ea
,Cafarella:2007pc
,Moretti:2001zz
,Kilian:2007gr
,Gleisberg:2008ta
,Kleiss:2010hy
,Kolodziej:2009uc
}.
By now, developments are heading at reaching this status to next-to-leading order in perturbation theory.
This includes one-loop amplitudes and real-radiation contributions with all the complications arising due to mass singularities and the highly non-trivial phase space integration~\cite{%
 Berger:2008sj
,Cascioli:2011va
,Cullen:2011ac
,Badger:2012pg
,Hirschi:2011pa
,Bevilacqua:2011xh
,Actis:2012qn
}.

HEF requires partonic scattering amplitudes with off-shell initial state partons, and automation of the calculation of these has not been established.
Systematic formulations of their calculation have been established~\cite{Lipatov:1995pn,Lipatov:2000se,Antonov:2004hh,vanHameren:2012if,vanHameren:2013csa}.
In this paper, we present a program to numerically evaluate tree-level multi-gluon scattering amplitudes with up to two off-shell gluons as function of the gluon momenta, squared and summed over colors of all gluons, and summed over the spins of the on-shell gluons.
It evaluates them via color-ordered helicity amplitudes that are calculated using the generalization of Britto-Cachazo-Feng-Witten (BCFW) recursion~\cite{Britto:2004ap,Britto:2005fq}, described in~\cite{vanHameren:2014iua}, to include off-shell gluons.
The program uses both hard-coded expression obtained via analytical BCFW recursion, and numerical BCFW recursion.
Using the latter, color-ordered amplitudes may be calculated to essentially arbitrary multiplicity.
Squared amplitudes summed over colors and helicities are provided for up to six external gluons.

This paper continues as follows: in \Section{Sec:definitions} the amplitudes that the program calculates are defined.
\Section{Sec:color} explains how color is treated.
\Section{Sec:usage} describes the usage of the program, and \Section{Sec:validation} introduces the program with which it was validated.
\Section{Sec:summary}, finally, contains the summary.

%% file: 2_definitions.tex
\section{Definitions\label{Sec:definitions}}
We consider the generally factorized formula for the gluonic contribution to a cross section in hadron collisions
%
\begin{equation}
\sigma\big(h_1(p_1)h_2(p_2)\to X\big) = \int d^4k_1\,F_1(k_1) \int d^4k_2\,F_2(k_2)\,
\frac{\hat{\sigma}\big(g^*(k_1)g^*(k_2)\to X\big)}{4\sqrt{(\lop{k_1}{k_2})^2-k_1^2k_2^2}}
~.
\end{equation}
%
This formula is very general, and $g^*$ does not necessarily refer to an off-shell gluon.
In collinear factorization, for example, we would have
%
\begin{equation}
F_i(k_i) = \frac{1}{2N_c}\int_0^1\frac{dx_i}{x_i}\,f_i(x_i,\mu)\,\delta^4(k_i-x_i\,p_i)
~,
\end{equation}
%
where $f_i$ is the collinear pdf for a hadron of type $i$.
We include the factors establishing averaging over spins and colors in $F_i$ here.
In the hybrid HEF~\cite{Deak:2009xt}, for example, $F_2$ would be as above, while $F_1$ would be given by
%
\begin{equation}
F_1(k_1) =
\frac{1}{N_c}\int \frac{d^2k_T}{2\pi}\int_0^1\frac{dx_1}{x_1}\,\EuScript{F}_1(x_1,k_T,\mu)
\,\delta^4(k_1-x_1\,p_1-k_T)
~,
\end{equation}
%
where $\EuScript{F}_1(x_1,k_T,\mu)$ is the unintegrated gluon density.

The symbol $X$ stands for a partonic final state, for example a number of on-shell gluons.
The partonic cross section $\hat{\sigma}$ is given by
%
\begin{equation}
\hat{\sigma}\big(g^*(k_1)g^*(k_2)\to X\big)
=
\int d\Phi(k_1,k_2\to X)\,\big|\Amp(g^*g^*\to X)\big|^2\,\EuScript{O}(X)
~.
\end{equation}
%
The phase space integration includes the summation over all color and spin degrees of freedom.
The observable $\EuScript{O}$ turns the partonic final state into a physical final state, for example through a jet algorithm.
This function also contains the necessary symmetry factors related to the final state.
We concentrate on the amplitude $\Amp(g^*g^*\to X)$ from now on, for the case that $X$ stands for a number of on-shell gluons.

We will adopt the convention that momenta denoted by the letter $p$ are always light-like, while momenta denoted by the letter $k$ are not necessarily light-like.
For the multi-gluon amplitudes we consider, this means that the initial-state momenta are denoted $k_1^\mu,k_2^\mu$, while the final-state momenta are denoted $p_3^\mu,\ldots,p_n^\mu$.
Momentum conservation is imposed as
%
\begin{equation}
k_1^\mu+k_2^\mu+p_3^\mu+\cdots+p_n^\mu = 0
~,
\end{equation}
%
so the initial-state momenta have negative energy.

Spin amplitudes with on-shell gluons depend on the momenta $p_i^\mu$ and polarization vectors $\vep_i^\mu$ associated with those gluons.
On-shellness implies that for each on-shell gluon we have
%
\begin{equation}
\lop{p_i}{p_i}=0
\quad\textrm{and}\quad
\lop{p_i}{\vep_i}=0
~.
\end{equation}
%
Gauge invariance assures the Ward identity that any momentum proportional to $p_i^\mu$ may be added to $\vep_i^\mu$ without changing the amplitude:
%
\begin{equation}
\Amp(\ldots;\,p_i^\mu,\vep_i^\mu\,;\ldots)
=
\Amp(\ldots;\,p_i^\mu,\vep_i^\mu+z\,p_i^\mu\,;\ldots)
\quad\forall\,{z\in\mathbf{C}}
~.
\end{equation}
%
Consequently, the amplitude only depends on the transverse part of $\vep_i^\mu$, and one may consider helicity amplitudes instead, which depend on $p_i^\mu$ and the helicity $\lambda_i$ which takes the two possible values $+$ or $-$.

Regarding off-shell gluons, the amplitude depends on their momenta $k_i^\mu$, and on their ``polarization vector'' or {\em direction\/} $p_i^\mu$, satisfying
%
\begin{equation}
\lop{p_i}{p_i}=0
\quad\textrm{and}\quad
\lop{p_i}{k_i} = 0
~.
\label{eq:direction}
\end{equation}
%
Within HEF, the direction is given by the momentum of one of the scattering hadrons, and the off-shell momentum is typically defined in terms of this and a transverse momentum via
%
\begin{equation}
k_i^\mu = x_i\,p_i^\mu + k_{T,i}^\mu
~.
\end{equation}
%
Realize, however that given any momentum $k_i^\mu$, one may construct an associated direction $p_i^\mu$ satisfying \Equation{eq:direction} (see \Appendix{App:A}).
For the amplitude, the notion of transverse momentum is arbitrary to a certain degree, since one can shift a fraction of $p_i^\mu$ to $k_{T,i}^\mu$:
%
\begin{equation}
k_i^\mu = x_i'\,p_i^\mu + k_{T,i}'^\mu
\quad\textrm{with}\quad
x_i' = x_i-x
\quad,\quad
k_{T,i}'^\mu = k_{T,i}^\mu + x\,p_i^\mu
~.
\end{equation}
%
Interpreting this as a change of $p_i^\mu$ rather than $x_i$, the amplitude scales homogeneously with this change:
%
\begin{equation}
\Amp\left(\ldots;\,k_i^\mu,\frac{x_i'}{x_i}\,p_i^\mu\,;\ldots\right)
=
\frac{x_i'}{x_i}\,\Amp(\ldots;\,k_i^\mu,p_i^\mu\,;\ldots)
~.
\end{equation}
%

%% file: 3_color.tex
\section{Color treatment\label{Sec:color}}
One issue regarding the color treatment as presented in~\cite{vanHameren:2012if} has to be settled.
There, the amplitude with off-shell gluons is obtained by considering each of them as an auxiliary eikonal quark-anti-quark pair, carrying fundamental color indices.
This situation is different from~\cite{Kotko:2014aba}, where each off-shell gluon carries a single adjoint color index, and the well-known color decompositions, the one presented in~\cite{DelDuca:1999rs} in particular, hold manifestly.
It is not {\it a priori\/} clear that these also hold in the formulation of~\cite{vanHameren:2012if}.
In particular, one might expect that the so-called $U(1)$-gluons would contribute in the color-flow representation, connecting two eikonal quark lines.
In this section, we will argue that this is not the case, and that the representations of~\cite{Kotko:2014aba} and~\cite{vanHameren:2012if} are indeed equivalent.

This is essentially guaranteed by the (proven) observation, stated in eq.(40) and eq.(41) of~\cite{vanHameren:2012if}, that the {\em induced vertices\/} of figure~{4} in~\cite{vanHameren:2012if} are traceless.
This also guarantees the equivalence of these vertices with those defined in~\cite{Antonov:2004hh}.
Consider an amplitude in the color representation in which each gluon, be it on-shell or off-shell, is represented by a pair of fundamental color indices.
For the on-shell gluons this is achieved by contracting the amplitude with $T^{a_l}_{j_li_l}$, where $a_l$ is the adjoint color index of the external on-shell gluon.
Tracelessness of the induced vertices implies tracelessness with respect to the color indices of the auxiliary eikonal quark-anti-quark pair, so for each pair of fundamental color indices referring to a gluon in the amplitude, on-shell and off-shell, we have
%
\begin{equation}
\EuScript{M}^{i_1\cdots i_n}_{j_1\cdots j_n}\,\delta^{j_g}_{i_g} = 0
~.
\label{eq:traceless}
\end{equation}
%
Each pair $i_l,j_l$ may refer to a gluon, on-shell or off-shell, or an ordinary quark-anti-quark pair.
The relation above only holds if $g$ refers to a gluon.
The general formula for the color-flow decomposition of the amplitude is given by~\cite{Kanaki:2000ms,Papadopoulos:2005ky}
%
\begin{equation}
\EuScript{M}^{i_1\cdots i_n}_{j_1\cdots j_n}
=
\sum_{\sigma\in S_n} \delta^{i_1}_{j_{\sigma(1)}}\cdots\delta^{i_n}_{j_{\sigma(n)}} \Amp'_\sigma
~,
\end{equation}
%
where the sum is over {\em all\/} $n!$ permutations $\sigma$ of $(1,2,\ldots,n)$.
Thanks to \Equation{eq:traceless}, all partial amplitudes $\Amp'_\sigma$ vanish if $l=\sigma(l)$ for any $l$ refering to a gluon.
As a result, for the case there are only gluons, the decomposition reduces to~\cite{Maltoni:2002mq}
%
\begin{equation}
\EuScript{M}^{i_1\cdots i_n}_{j_1\cdots j_n}
=
\sum_{\sigma\in S_{n-1}} \delta^{i_{\sigma(1)}}_{j_{\sigma(2)}}\,\delta^{i_{\sigma(2)}}_{j_{\sigma(3)}}\cdots\delta^{i_{\sigma(n-1)}}_{j_{n}}\,\delta^{i_{n}}_{j_{\sigma(1)}} \Amp_\sigma
~,
\end{equation}
%
where the sum is now over only $(n-1)!$ permutations.
The change in notation from $\Amp'_\sigma$ to $\Amp_\sigma$ just indicates that the labelling is not identical, \eg\ $\Amp_{1234}=\Amp'_{2341}$ etc..
We may return to the adjoint representation by contracting every fundamental pair $i,j$ with $T^a_{j,i}$, leading to the well-known formula~\cite{Berends:1987cv,Mangano:1987xk}
%
\begin{equation}
\EuScript{M}^{a_1\cdots a_n}
=
\sum_{\sigma\in S_{n-1}} \Tr\left\{T^{a_{\sigma(1)}}T^{a_{\sigma(2)}}\cdots T^{a_{\sigma(n-1)}}T^{a_n}\right\}\,\Amp_\sigma
~.
\end{equation}
%
The formulas above are not quite the same as in the mentioned references.
One ingredient that is missing is that, if all gluons are on-shell, the partial amplitudes $\Amp_\sigma$ are given by a single amplitude with permuted arguments:
%
\begin{equation}
\Amp_\sigma(1,\ldots,n-1,n) = \Amp(\sigma(1),\ldots,\sigma(n-1),n)
\label{eq:permargs}
~.
\end{equation}
%
Each argument in the form of a number, say $l$, refers to both the momentum and the helicity of gluon $l$.
This is easy to understand considering that all external gluons are essentially equivalent.
If some of them are off-shell, this does not seem so obvious, but we may consider the case in which {\em all of them\/} are off-shell, and thus equivalent.
Now, each number $l$ refers to the momentum and direction, or longitudinal momentum component, associated with off-shell gluon $l$.
We may take the on-shell limit of gluon $l$, which in~\cite{vanHameren:2014iua} is argued to lead to
%
\begin{align}
\EuScript{M}^{a_1\cdots a_n}
&\to&
\sum_{\sigma\in S_{n-1}} \Tr\left\{T^{a_{\sigma(1)}}T^{a_{\sigma(2)}}\cdots T^{a_{\sigma(n-1)}}T^{a_n}\right\}\ \Amp(\sigma(1),\ldots,\sigma(l^+),\ldots,\sigma(n-1),n)
\nonumber\\
&+&
\sum_{\sigma\in S_{n-1}} \Tr\left\{T^{a_{\sigma(1)}}T^{a_{\sigma(2)}}\cdots T^{a_{\sigma(n-1)}}T^{a_n}\right\}\ \Amp(\sigma(1),\ldots,\sigma(l^-),\ldots,\sigma(n-1),n)
\end{align}
%
It is explained in~\cite{vanHameren:2014iua} how this coherent sum over helicities turns into an incoherent sum for the squared amplitude.
We see that for each helicity, the color decompostion including \Equation{eq:permargs}, still holds.
The color decompostion presented in~\cite{DelDuca:1999rs} also holds, since it is proven there using only group theory arguments, and is independent of the exact form of the partial amplitudes.

%% file: 4_usage.tex
\newpage
\section{Usage\label{Sec:usage}}
The program can be obtained from
\begin{lstlisting}
  http://bitbucket.org/hameren/amp4hef
\end{lstlisting}
The main directory contains a \codetxt{README}\ file describing how to compile and use the program.
There is an example directory with a simple Monte Carlo program to show how to use the program explicitly.
The program is written in Fortran~2003, but it eventually provides a module that makes a few subroutines available that only take (arrays of) intrinsic integer, real and complex type variables as arguments. 
More specifically, the
\begin{lstlisting}
  module amp4hef
\end{lstlisting}
provides first all the
\begin{lstlisting}
  subroutine init_amp4hef
\end{lstlisting}
which does not take any arguments, and must be called once before any  
other routine. 
Furthermore, it provides the
\begin{lstlisting}
  integer,parameter :: fltknd=kind(1d0)
\end{lstlisting}
which is set to this value in the \codetxt{module amp4hef\_spinors}, which can be found in the source file \codetxt{amp4hef\_spinors.f90}.
All real and complex variables are of this kind.
Then, the module provides
\begin{lstlisting}
  subroutine construct_ng( Noffshel ,Nonshell ,momenta ,directions )
    integer     ,intent(in) :: Noffshell ,Nonshell
    real(fltknd),intent(in) :: momenta(0:3,*) ,directions(0:3,*)
\end{lstlisting}
This routine takes external momenta and directions as arguments, and prepares all spinors etc.\ for the evaluation of amplitudes and matrix elements.  
The input \codetxt{Noffshell} refers to the number of off-shell momenta, and should be $0$, $1$ or $2$.
The input \codetxt{Nonshell}  refers to the number of on-shell momenta.
All momenta should be provided via the array \codetxt{momenta}.
The size of the second dimension  should be at least $n$$=$\codetxt{Noffshell}$+$\codetxt{Nonshell}.
The initial-state momenta should be \codetxt{momenta(}$\mu$\codetxt{,}$i$\codetxt{)}$=$$k_i^\mu$ for $i$$\in$$\{1,2\}$, while the final-state momenta should be \codetxt{momenta(}$\mu$\codetxt{,}$i$\codetxt{)}$=$$p_i^\mu$ for $i$$\in$$\{3,\ldots,n\}$, where $\mu$$\in$$\{0,1,2,3\}$.
The size of the second dimension of the input \codetxt{directions} should be at least \codetxt{Noffshell}.
This array should provide the directions associated with the off-shell momenta: \codetxt{directions(}$\mu$\codetxt{,}$i$\codetxt{)}$=$$p_i^\mu$.
They should satisfy \Equation{eq:direction}.

After calling \codetxt{subroutine construct\_ng}, the squared amplitude summed over colors and helicities can be evaluated for the given momenta and directions with
\begin{lstlisting}
  subroutine matrix_element_ng( rslt )
    real(fltknd),intent(out) :: rslt
\end{lstlisting}
The output of this routine is missing a factor $(4\pi\alpha_S)^{n-2}$, where $n$ is the total number of gluons.
The number of colors is fixed to $N_c=3$.
This can be changed in the file \codetxt{amp4hef\_ng.f90}.
The output of this routine is normalized such that when $k_i^\mu\to p_i^\mu$ for each off-shell gluon, then the result becomes the standard result with on-shell gluons only, including the sum over all their helicities.

Also after calling \codetxt{subroutine construct\_ng}, color-ordered helicity amplitudes can be evaluated with
\begin{lstlisting}
  subroutine amplitude_ng( rslt ,helicity ,permutation )
    complex(fltknd),intent(out) :: rslt
    integer        ,intent(in) :: helicity(*)
    integer        ,intent(in) :: permutation(*)
\end{lstlisting}
The size of the array \codetxt{helicities} should be at least $n$$=$\codetxt{Noffshell}$+$\codetxt{Nonshell}.
Its entries refer to the helicities of the on-shell gluons and follow the same enumeration as the momenta given to \codetxt{subroutine construct\_ng}.
Their values should be one of $-1$ or $+1$.
The size of the array \codetxt{permutation} should be at least $n-2$, and only refers to the final-state gluons.
Denoting $\lambda_i$$=$\codetxt{helicities(}$i$\codetxt{)} for $i$$\in$$\{1,2,\ldots,n\}$, and
$\sigma(i)$$=$\codetxt{permutation(}$i$\codetxt{)} for $i$$\in$$\{1,2,\ldots,n-2\}$, this routine returns the value of
%
\begin{equation}
\Amp\big(1,2\,
;\,p_{2+\sigma(1)}\,,\,\lambda_{2+\sigma(1)}\,
;\,p_{2+\sigma(2)}\,,\,\lambda_{2+\sigma(2)}\,
;\ldots
;\,p_{2+\sigma(n-2)}\,,\,\lambda_{2+\sigma(n-2)}\,
\big)
~.
\end{equation}
%
Besides the square root of the factor mentioned before, the output of this routine is also missing a factor $\sqrt{|k_i^2|}$ for each off-shell gluon.

Similarly to the routines described above, the \codetxt{module amp4hef} also  provides  
\begin{lstlisting}
  subroutine construct_ng_xpr(Noffshel,Nonshell,momenta,directions)  
  subroutine matrix_element_ng_xpr(rslt)  
\end{lstlisting}
which are defined in the file \codetxt{amp4hef\_ng\_xpr.f90}, and which use analytic expressions rather than numerical BCFW recursion.
The total number of gluons is restricted to a maximum of 5 for these.  
To give an impression of the size of the expressions, we present here one of them:
\begin{multline}
\Amp(1^*,2^*,3^-,4^-,5^+)=
 \frac{1}{\kapp_1\kapp_2}\frac{\SQR{12}^3\ANG{43}^3}
 {\AxS{5}{\slashk_1\!+\!\slashk_2}{2}\AxS{3}{\slashk_1\!+\!\slashk_2}{1}\ANG{54}(\slashk_1\!+\!\slashk_2)^2}
\\
+\frac{1}{\kapp_1\kstr_2}\frac{\ANG{32}^3\SQR{51}^3}
  {\AxS{2}{\slashk_2\!+\!\slashp_3}{4}\AxS{3}{\slashk_2}{1}(\slashk_2\!+\!\slashp_3)^2}
+\frac{1}{\kstr_1}\frac{\SQR{25}^4\ANG{21}^3}
  {\AxS{1}{\slashk_1\!+\!\slashk_2}{2}\AxS{2}{\slashk_1}{5}\AxS{2}{\slashk_1}{2}\SQR{23}\SQR{34}\SQR{45}}
\\
-\frac{1}{\kapp_2}\frac{\AxS{1}{\slashp_3\!+\!\slashp_4}{2}^4\SQR{12}^3}
  {\AxS{1}{\slashk_1\!+\!\slashk_2}{2}\SxS{2}{(\slashp_3\!+\!\slashp_4)(\slashk_1\!+\!\slashp_5)}{1}\AxS{5}{\slashp_3\!+\!\slashp_4}{2}\ANG{15}\SQR{23}\SQR{34}\big(\AxS{1}{\slashk_2}{1}\SQR{24}+\ANG{13}\SQR{34}\SQR{21}\big) }
\\
+\frac{1}{\kapp_1}\frac{\SQR{51}^3\AxS{2}{\slashp_4\!+\!\slashp_3}{2}^3}
  {\AxS{2}{\slashk_2\!+\!\slashp_3}{4}\AxS{2}{\slashk_1}{5}\SxS{2}{(\slashp_4\!+\!\slashp_3)(\slashk_1\!+\!\slashp_5)}{1}\SQR{23}\SQR{34}(\slashk_1\!+\!\slashp_5)^2}
\\
-\frac{1}{\kapp_2}\frac{\SQR{12}^3\ANG{13}^4\SQR{51}^3}
  {\AxS{3}{\slashk_2}{1}\AxS{1}{\slashk_2}{1}\AxS{3}{\slashk_1\!+\!\slashk_2}{1}\AxS{1}{\slashk_2\!+\!\slashp_3}{1}\SQR{45}\big(\AxS{1}{\slashk_2}{1}\SQR{24}+\ANG{13}\SQR{34}\SQR{21}\big)}
\end{multline}
When both off-shell gluons go on-shell, the first term corresponds to the helicities $(1^+,2^+)$, the second to $(1^+,2^-)$, while the third and fourth combine to $(1^-,2^+)$.
The last two terms are irrrelevant in that limit, and the helicities $(1^-,2^-)$ correspond to a vanishing amplitude in the on-shell case.

The \codetxt{module amp4hef} also provides the routines defined in the file \codetxt{amp4hef\_ons\_xpr.f90}, which collects some expressions from literature for squared amplitudes summed over colors and helicities~\cite{Kuijf:1991kn}.
These are mainly included to clarify the normalization conventions used in this program.

\section{Validation\label{Sec:validation}}
The Monte Carlo program in the example directory simulates the generation of events by reading event files that were produced with the help of A Very Handy LIBrary\footnote{{\tt http://bitbucket.org/hameren/avhlib}} in Fortran.
The files include the event weights obtained with AVHLIB, and have been used to validate the squared amplitudes summed over spins and color provided by AMP4HEF.
AVHLIB provides a completely independent implementation for the numerical evaluation of scattering amplitudes, employing Dyson-Schwinger recursion, for arbitrary processes, both with on-shell and off-shell initial-state partons, and including both QCD and electro-weak interactions.
Furthermore, it provides an efficient phase space generator~\cite{vanHameren:2007pt,vanHameren:2010gg} and other tools for the evaluation of distributions of arbitrary observables using the Monte Carlo method of integration, with which the aforementioned event files were produced.
The amplitudes can be provided in several representations by AVHLIB, among which is the decomposition into color-ordered helicity amplitudes.
So besides the squared amplitudes, also the color-ordered helicity amplitudes from AMP4HEF were validated seperately with the help of AVHLIB.

%% file: 8_references.tex
\providecommand{\href}[2]{#2}\begingroup\raggedright\endgroup

%% file: 9_appendixA.tex
\section{Construction of the direction\label{App:A}}
Given the Sudakov decomposition in terms of two light-like momenta $p_1^\mu,p_2^\mu$
%
\begin{equation}
k^\mu = x_1\,p_1^\mu + x_2\,p_2^\mu + k_T^\mu
\quad\textrm{with}\quad
\lop{p_1}{k_T} = \lop{p_2}{k_T} = 0
\quad\textrm{and}\quad
\lop{p_1}{p_2}=\srac{1}{2}\,s\neq0
~,
\end{equation}
%
we may choose
%
\begin{equation}
p^\mu = p_1^\mu - \frac{x^2k_T^2}{s}\,p_2^\mu - x\,k_T^\mu
\quad\textrm{with}\quad
x = \frac{1}{x_1}\left(\sqrt{1+\frac{x_1x_2s}{k_T^2}}-1\right)
~.
\end{equation}
%
We then have
%
\begin{equation}
k^\mu = x_1p^\mu + k_T'^{\mu}
\end{equation}
%
with $p^\mu$ defined above, and with
%
\begin{equation}
k_T'^{\mu} = (1+xx_1)\,k_T^\mu + \left(\frac{x^2k_T^2}{s}\,x_1+x_2\right)p_2^\mu
~,
\end{equation}
%
satisfying
%
\begin{equation}
\lop{p}{p}=\lop{p}{k_T'}=0
~.
\end{equation}
%
Notice that $x\to0$ for $x_2\to0$, so the construction is consistent.
Also, if $k^\mu$ becomes light-like, that is if $x_1x_2s+k_T^2\to0$, we have $x\to-1/x_1$ and $k_T'^\mu\to0$.

Alternatively, if $k^\mu$ is given as the sum of two light-like momenta,
%
\begin{equation}
k^\mu = p_1^\mu + p_2^\mu
\label{eq:lldecom}
~,
\end{equation}
%
we may choose
%
\begin{equation}
p^\mu = p_1^\mu - p_2^\mu
      - \frac{1}{2z}\,\AxS{p_1}{\gamma^\mu}{p_2}
      + \frac{z}{2}\,\AxS{p_2}{\gamma^\mu}{p_1}
\end{equation}
%
for any $z\neq0$, or in terms of spinors:
%
\begin{equation}
\RA{p} = \RA{p_1} + z\,\RA{p_2}
\quad,\quad
\RS{p} = \RS{p_1} - \frac{1}{z}\,\RS{p_2}
~.
\end{equation}
%
A decomposition like \Equation{eq:lldecom} can always be constructed given any auxiliary light-like momentum $q^\mu$ with $\lop{q}{k}\neq0$:
%
\begin{equation}
p_2^\mu = \frac{k^2}{2\lop{q}{k}}\,q^\mu
\quad,\quad
p_1^\mu = k^\mu-p_2^\mu
~.
\end{equation}
%

%
%